\documentclass[aps,twocolumn, amsmath,amssymb, floatfix, citeautoscript]{revtex4}

\usepackage{graphicx, subfigure}
\usepackage{hyperref}
\usepackage{dcolumn}
\usepackage{bm}

\begin{document}


\title{A numerical procedure for model reduction using the generalized Langevin equation formalism}
\author{Abhishek Mukhopadhyay}
\affiliation{Departments of Physics and Computer Science, Virginia Polytechnic Institute and State University, Blacksburg, VA 24061}
\author{Jianhua Xing}
\email{jxing@vt.edu}
\affiliation{%
Department of Biological Sciences, Virginia Polytechnic Institute and State University, Blacksburg, VA 24061}
\date{\today}

\begin{abstract}
The Zwanzig-Mori projection formalism is widely used in studying  systems with many degrees of freedom. Recently Xing and Kim used the projection formalism and derived the generalized Langevin equations (GLEs) for a general stochastic system not necessarily obeying detailed balance. In this study we develop a numerical procedure to  reconstruct the GLEs from data. Numerical tests on two biological networks show remarkable agreement between the results calculated from the reconstructed GLEs and those of full model simulations. We suggest that the procedure can be applied in model reduction and a novel way of nonlinear time series analysis.
\end{abstract}

\pacs{Valid PACS appear here}
\maketitle

\section{\label{sec:Introduction}Introduction}
It is common to study dynamics of a system with  a large number of degrees of freedom in almost every scientific field. In general it is impractical, and often unnecessary,  to track all the dynamic information of the whole system. Furthermore even if the dynamic information of the whole system is available,  a set of equations of motion or representations  at reduced dimension are often desirable to reveal the system dynamics more transparently and informatively, or to allow efficient description of long time dynamics. Consider, for example, a protein may have tens of thousands of atoms interacting with an even larger number of solvent molecules and other molecules. However, to provide insights of the protein functional mechanism it is often needed to monitor a few number of collective modes.
 
 In statistical physics, the celebrated Zwanzig-Mori projection approach is a general and formal procedure to derive the equations of motion of a set of selected primary degrees of freedom,   while  the remaining secondary degrees of freedom manifest their effects on the primary degrees implicitly.
 This basic idea can be dated back to Einstein on his treatment on Brownian motion~\cite{Einstein1905}. Inspired by techniques from quantum mechanics, Zwanzig provided a formal procedure leading to a set of  governing equations in the form of generalized Langevin equations (GLEs)~\cite{Zwanzig1960,Zwanzig1961}. In the limit that there is clear time scale separation between the primary and secondary degrees of freedom, the GLEs can reduce into Langevin equations. Mori further derived a simplified version of the equations that is "inherently linear in the system variables"~\cite{Mori1965, Zwanzig2001}. The Z-M procedure is typically adopted to studies on closed Hamiltonian systems that relax to thermodynamic equilibrium in the long time limit. One notable recent development is the work of Lange and Grubmuller to derive the dynamic equations of some collective coordinates with the Zwanzig projection procedure~\cite{Lange2006}. 

The success of the Zwanzig-Mori projection approach leads to great interest to apply it to general non-Hamiltonian systems out of thermal dynamic equilibrium, which may be dynamically highly inhomogeneous, and have fascinating emerging properties such as oscillations and bifurcations. Examples include climate changes in one place, stock market fluctuations, and collective dynamics of $\sim 10^{11}$ neurons in a brain, stochastic dynamics of gene regulatory networks. It is of both theoretical and practical significances to examine the applicability of the projection formalism to complex systems, if yes, what new features of the GLEs exist.    
Chorin and coworkers have made considerable efforts in this field~\cite{Chorin2000, Chorin2002}. Using a time series analysis approach\cite{Erban2006}, Erban et al. showed how one can reconstruct a one-dimensional Fokker-Planck equation, or the equivalent Langevin equation, to recapitulate the dynamics of a gene regulatory network such as a toggle switch.  Kawai and Komatsuzaki derived the GLEs in nonstationary environments~\cite{Kawai2011}. 

Recently Xing and Kim applied the projection approach to nonequilbrium non-Hamiltonian systems with stochastic dynamics and derived the corresponding GLEs. In this paper we will further present a numerical procedure on how to reconstruct GLEs in a reduced representation from full model simulations. A prominent feature of the procedure is that all the model parameters are extracted directly from the data, in the same spirit of Erban {\it et al}\cite{Erban2006}. The remaining part of the paper is organized as follows: in Sec. \ref{sec:Theory} we summarize the theory of the Zwanzig-Mori projection and its generalization to non-Hamiltonian systems; in Sec. \ref{sec:Methods} we present the numerical approach to extract the GLE model parameters; Sec. \ref{sec:Results} shows numerical tests on two model systems; and we conclude with discussions in Sec. \ref{sec:Discussions}. 

\section{\label{sec:Theory}Theory}
First we  give a heuristic presentation on the projection approach, following the discussions given in~\cite{Zwanzig2001} with some modifications, and focusing on the Hamiltonian systems.

Consider a system described by the Hamiltonian,
\begin{eqnarray}
H(\mathbf{x,p}) = \sum_{i=1}^n\frac{p_i^2}{2}+V(\mathbf{x})
\end{eqnarray}
where $\mathbf{x}$ and $\mathbf{p}$ are position and conjugate momentum vectors. We will use mass-weighted coordinates here.

The Liouville operator $L$ is defined as,
\begin{eqnarray}
LA = \sum_i \left( \frac{\partial H}{\partial p_i}\frac{\partial A}{\partial x_i} - \frac{\partial H}{\partial x_i}\frac{\partial A}{\partial p_i}  \right)
\end{eqnarray} 
For an arbitrary dynamic variable $A$,  the projection operator  is defined as,
\begin{eqnarray}
PA  &=&  \sum_{ij} (A, \phi_i)(\phi, \phi)^{-1}_{ij}\phi_j,
\end{eqnarray}
where $\{\phi(\mathbf{x},\mathbf{p} )\}$ composes the basis set for the projected subspace. The inner product  for two arbitrary variables $A$ and $B$ is defined as,
\begin{eqnarray}
(A,B) &=& <A^\dagger B>\nonumber\\
&=&\frac{\int   A^\dagger B \exp\left( -\beta H \right) d \mathbf{x} d \mathbf{p} }{\int  \exp\left( -\beta H \right) d \mathbf{x} d \mathbf{p} } \nonumber
\end{eqnarray}
where $\dagger$ means taking transpose and complex conjugate. Any dynamic variable within the subspace can be expressed as a linear combination of the basis functions. The projected equations of an arbitrary dynamic variable $A$, which is defined within the projected subspace, are given by,
\begin{eqnarray}
\frac{\partial}{\partial t}A(t)& =& PLA(t) -\int_0^t ds\mathbf{K}(s)\cdot \phi(\mathbf{x}(t-s),\mathbf{p}(t-s) \nonumber\\
&& + F(t) \label{eqn:GLE_formal} 
\end{eqnarray}
where 
\begin{eqnarray}
F(t)  &=&  \exp(t(\mathbf{1-P})L)(\mathbf{1-P})LA\\
\mathbf{K}(t) &=& -(LF(t),\phi) \cdot (\phi,\phi)^{-1} \nonumber\\
&=& (F(t),L\phi)\cdot (\phi,\phi)^{-1} \label{eqn:memory_general}
\end{eqnarray}
The last equation leads to the generalized fluctuation-dissipation relation, and we have used the anti-Hermitian property of the Liouville operator.

In practice the basis sets are usually chosen as portion of the coordinate vector $\mathbf{x}$   and the corresponding conjugate momentum vector $\mathbf{p}$. Mori derived a GLE that is linear to the coordinates and momentum~\cite{Mori1965}. However, in principle this restriction is unnecessary. One can expand the Hilbert space
to include high order combinations of the coordinates and momentum. Appendix \ref{appendixAnalytic} gives such an analytic example. Therefore it should be clear that while the projection is performed in the linear Hilbert space, the resultant GLEs can be nonlinear to the coordinate variable.
Especially below let's consider  the extreme limit of including all the possible Hilbert functions composed by the coordinate and velocity (or momentum) in reduced dimension.  The following procedure is analogous to what adopted by Zwanzig~\cite{Zwanzig1961}. 

For simplicity let's focus on projecting onto a one-dimensional manifold $c$ and its conjugate momentum, while generalization to higher dimensions is straightforward. Noticing that a possible choice of  the basis set of the Hilbert space is  $\{\mathbf{x}, \mathbf{\dot{x}}, \mathbf{xx}, \cdots \}$, one can expand $c$ and ${\dot {c}}$ as 
 \begin{eqnarray}
 c &=& f(\mathbf{x})= f(0) + \nabla_{\mathbf{x}}f(0)\cdot \mathbf{x} + \frac{1}{2}\nabla_{\mathbf{xx}}f(0)\colon \mathbf{xx} +\dots\\
 \dot{c} &=&  \nabla_{\mathbf{x}} f \cdot \mathbf{\dot{x}}= \nabla_{\mathbf{x}} f(0) \cdot \mathbf{\dot{x}}
    + \nabla_{\mathbf{xx}} f(0) \colon \mathbf{x\dot{x}}+\dots
 \end{eqnarray}
Therefore $c$ and $\dot{c}$ are vectors in the full Hilbert space. Let's consider the sub-Hilbert space, which may still have infinite dimension, supported by  all the possible multiplicative combinations of $c$ and $\dot{c}$ , such as $c^2, \dot{c} c^3\dots$. 
 A key observation is that these basis functions, denoted $\{\phi(c,\dot{c})\}$, compose a complete basis set for the subspace so any arbitrary function of $(c,\dot{c})$,  can be fully expressed by the basis set. That is, for an arbitrary function $g(\mathbf{x,p})$, the inner product becomes
 \begin{eqnarray}
&& \sum_i\int g \phi_i \exp(-\beta H) d\mathbf{x}d\mathbf{p} \nonumber\\
 && =\frac{1}{ \bar{\rho}(c,\dot{c})}
  \int g  \exp(-\beta H) \delta(f-c)\delta(\nabla_\mathbf{x} f \cdot \mathbf{p} - \dot{c}) d\mathbf{x}d\mathbf{p} \nonumber
 \end{eqnarray} 
 where
 \begin{eqnarray}
 \bar{\rho}(c,\dot{c}) = {\int   \exp(-\beta H) \delta(f-c)\delta(\nabla_\mathbf{x}f \cdot \mathbf{p} - \dot{c}) d\mathbf{x}d\mathbf{p} }
 \end{eqnarray}
The above expression may be more familiar if the Dirac bra and ket notations are used. Then one obtains explicit expressions for the first term on the right hand side of  Eqn. \ref{eqn:GLE_formal}, 
\begin{eqnarray}
PL \cdot c &=& \dot{c} \label{eqn:collective_c}\\
PL \cdot \dot{c} &=&  -k_BT\frac{1}{\bar{\rho}(c,\dot{c})}\frac{\partial}{\partial_c}  \int   \exp(-\beta H) \nonumber\\
      && ||\nabla_{\mathbf{x}} f ||^2 \delta(f-c)\delta(\nabla_\mathbf{x}f \cdot \mathbf{p} - \dot{c}) d\mathbf{x}d\mathbf{p}\label{eqn:collective_cdot} 
\end{eqnarray} 
To derive the above expression, we perform integration by parts, and use the relations,
\begin{eqnarray}
\nabla_{\mathbf{x}} \delta(c-f) &=&  \nabla_{\mathbf{x}}f \partial_f \delta(x-f) =   \nabla_{\mathbf{x}}f \partial_c\delta(x-f) \nonumber\\
\nabla_{\mathbf{x}} \delta(\dot{c}-\nabla_{\mathbf{x}}f \cdot \mathbf{p}) &=& \nabla_{\mathbf{x}} (\nabla_{\mathbf{x}}f \cdot \mathbf{p}) \partial_{\dot{c}} \delta(\dot{c}-\nabla_{\mathbf{x}}f \cdot \mathbf{p})  \nonumber\\
\nabla_{\mathbf{p}} \delta(\dot{c}-\nabla_{\mathbf{x}}f \cdot \mathbf{p}) &=& \nabla_{\mathbf{x}}f  \partial_{\dot{c}} \delta(\dot{c}-\nabla_{\mathbf{x}}f \cdot \mathbf{p})
\end{eqnarray}
Compared to Eqn. \ref{eqn:collective_cdot}, the result derived by Lange and Grubmuller has an extra term $||\nabla_{\mathbf{x}}f||^2$ in the expression of $\bar{\rho}(c, \dot{c} )$~\cite{Lange2006}. The discrepancy may come from the fact that the projection operator
defined in~\cite{Lange2006} does not rigorously satisfy $P^2=P$. It remains to be examined on how this extra term may affect the dynamics. In the case $f$ is a linear combination of $\mathbf{x}$, and is chosen to satisfy $||\nabla_{\mathbf{x}}f||^2=1$ , Eqn. \ref{eqn:collective_cdot}
gives the familiar relation to the gradient of potential of mean force.

Recently Xing and Kim\cite{Xing2011} applied the Zwanzig-Mori projection procedure to a general dynamic system described by a set of stochastic differential equations,
\begin{equation}
\dot{x}_{i}=G_i(\mathbf{x})+\sum_{j=1}^{M}g_{ij}(\mathbf{x})\xi_j(t)\mbox{  }i=1,\cdots,N
\label{eq:glegen}
\end{equation} 
In general the vector $\mathbf{G(x)}$ can not be represented as the gradient of a scalar potential due to violation of detailed balance, $M$ and $N$ may be different, $\xi_i(t)$ are temporally uncorrelated, statistically independent Gaussian white noise with the averages satisfying $\langle\xi_i(t)\xi_j(\tau)\rangle=\delta_{ij}\delta(t-\tau)$, $\mathbf{g(x)}$ is related to the $N\times N$ diffusion matrix $\mathbf{gg}^T = 2\mathbf{D}$, where the transpose of a matrix is designated by the superscript $T$. The resultant GLEs for projection to $\mathbf{X}$, components of $\mathbf{x}$, assume the form
\begin{eqnarray}
0&=& -\frac{\partial}{\partial X_i}W(\mathbf{X})-\sum_j \Gamma_{0,ij}\dot{X}_{j}(t)  \nonumber\\
&&  -\int_{0}^{t}ds\Gamma_{1,ij}(s)\dot{X}_j(t-s)+\mathcal{F}_{i}(t)
\label{eqn:GLE_phe}
\end{eqnarray}
where $W$ is the potential of mean force, $\Gamma_0$ is in general not symmetric, reflecting violation of detailed balance, and the generalized fluctuation-dissipation theorem reads,
\begin{equation}
<F_i(t)F_j(t')> = (\Gamma_{0,ij}+\Gamma_{0,ji})\delta(t) + \Gamma_{1,ij}(t-t')
\end{equation}
where $t\geq t'$. For simplicity in the remaining discussions we focus on the case that the projected system is one-dimensional, while generalization to higher dimensions is straightforward.

\section{\label{sec:Methods} Methods}
\subsection{Discretization of the Generalized Langevin Equation} 

In this work, we assume that the full dynamics of the system under study, or the time series data of the primary degrees of freedom  are available, and our goal is to reconstruct the dynamic equation in the reduced dimension from the data. 
 
Consider a generic 1-D  generalized Langevin equation, 
\begin{equation}
0 = \frac{\partial W}{\partial X(t)}+\left[\Gamma_0\dot{X}(t)+\int_{0}^{t}\Gamma_1(s)\dot{X}(t-s)ds\right]+\mathcal{F}(t)
\label{eq:glealt}
\end{equation} 
we first integrate it from time $i\Delta t$ to $(i+1)\Delta t$ for discrete time steps $\Delta t$, and perform ensemble average over the random force. Notice that the random force $\mathcal{F}(t)$ averages out to 0, then Eqn.\ref{eq:glealt} becomes,
\begin{eqnarray}
0&= &\int_{i\Delta t}^{(i+1)\Delta t}\langle\frac{\partial W(t)}{\partial X}\rangle \\ \nonumber
&&+\int_{i\Delta t}^{(i+1)\Delta t}\left[\langle\Gamma_0\dot{X}(t)\rangle+\langle\int_{0}^{t}\Gamma_1(s)\dot{X}(t-s)ds\rangle\right]
\label{eq:gleavg}
\end{eqnarray}

This can further be approximated as,
\begin{eqnarray}\label{eq:memk}
0&\simeq&\langle\frac{\partial W}{\partial X(t_i)}\rangle+\Bigg\{\frac{\Gamma_0}{\Delta t}\langle X(t_{i+1})-X(t_i)\rangle\\ \nonumber
&&+\sum_{j=1}^{i-1}\Gamma_1(t_{j+1/2})\langle(X(t_{i-j+1})-X(t_{i-j}))\rangle\Bigg\}
\end{eqnarray}

Alternatively, by multiplying $X(t_0)$ on both sides of Eqn.~\ref{eq:glealt} before performing average and integration, one obtains an equation for the autocorrelation function counterpart of Eqn.~\ref{eq:memk},

\begin{eqnarray}\label{eq:memcorr}
0 &   \simeq&  \langle X(t_0) \frac{\partial W}{\partial X(t_i)}\rangle  \\
           &&            +   \Bigg\{\frac{\Gamma_0}{\Delta t}\left[\langle X(t_0)X(t_{i+1})\rangle- \langle X(t_0)X(t_i)\rangle\right]    \nonumber\\
&&                 +\sum_{j=1}^{i-1}\Gamma_1(t_{j+1/2})  \left[  \langle X(t_0)X(t_{i-j+1})\rangle - \langle X(t_0) X(t_{i-j})\rangle \right] \Bigg\} \nonumber
\end{eqnarray}

The potential of mean force $W$ can be obtained directly from the stationary distribution $\rho_{ss(X)}$ from $W(X) = -\ln \rho_{ss}(X)$, and all the ensemble averaged terms ($\langle\cdot\rangle$) can be obtained from the data. 
\subsection{Smoothing the Memory Kernel with Tikhonov Regularization\label{subset:TR}}
Both Eqns.~\ref{eq:memk} and~\ref{eq:memcorr} are linear equations of the terms $\Gamma_0$ and $\Gamma_1(t)$. For example, Eqn.\ref{eq:memk} can be rewritten in matrix form as,
\begin{equation}
\mathbf{\partial_XW}-\Delta\mathbf{X\cdot}\Gamma= 0
\label{eq:tikho}
\end{equation}
Here $\mathbf{\partial_x W}_i=\langle\frac{\partial W}{\partial X(t_i)}\rangle$, elements of the square matrix 
$\Delta {X}_{ij}=\langle X(t_{i-j}) - X(t_{i-j+1})\rangle$ for $j\le i$ and 0 otherwise and 
$\Gamma_i=\Gamma(t_{i+1/2})$. 
 However, it is not numerically desirable to solve any of them directly, since numerical errors in $\mathbf{\partial_x W}_i$ and  $\Delta\mathbf{X}$ accumulate quickly and lead to inaccurate and unstable prediction of the GLE parameters. Instead we regularize the data 
using the Tikhonov regularization (also know as Ridge regularization)~\cite{Hastie2009}. Simply speaking, the Tikhonov regularization 
adds a  penalty term to damp the highly  oscillating components of the estimator.
Mathematically, we need to minimize the function;
\begin{equation}
||\mathbf{\partial_xW}-\Delta\mathbf{X\cdot}\Gamma||^2+||\mathbf{P\cdot}\Gamma||^2
\label{eq:tikho2}
\end{equation}
In this work we choose  $\mathbf{P_{ij}}=b$, for $i=j$ and $=-b$ for $i=j+1$ and 0 otherwise, except for the 
first and the last row where all elements are 0. the term $b$ is a preselected constant, which we will refer  as the penalization factor.
This minimization yields a simple solution of the form,
\begin{equation}
\Gamma=\left(\Delta\mathbf{X^T\cdot}\Delta \mathbf{X+P^TP}\right)^{-1}\Delta\mathbf{X^T\cdot\partial_xW}
\label{eq:tikho3}
\end{equation}
 We assume that $\Gamma_1$ is a smooth function of $t$, and thus the penalty function $\mathbf{P}$ is chosen to minimize the derivatives, or differences between $\Gamma_1$ at two consecutive time points. However, use of the function $\mathbf{P}$ may also lose  meaningful information and  distort the result. Hence the penalty factor, $b$,  must be chosen wisely. Below we demonstrate how to choose $b$ properly using two examples. 

\subsection{Generating random force $\mathcal{F}(t)$}
Using Eqn.~\ref{eq:glealt} and with $W$, $\Gamma_0$, and $\Gamma_1(t)$ determined, we simulate the GLE using the following discretized version,
\begin{eqnarray}
0&\simeq& -\frac{\partial W(X(t_{i-1}))}{\partial X}\Delta t-\Gamma_0\left(X(t_i)-X(t_{i-1})\right)\\ \nonumber
&-&\Delta t \sum_{k=0}^{i}\Gamma_1\left(\left(k+\frac{1}{2}\right)\Delta t\right)\left(X(t_{i-k})-X(t_{i-k-1})\right)\\ \nonumber
&+&\int_{t_{i-1}}^{t_i} dt \mathcal{F}(t)
\label{eq:dgle}
\end{eqnarray}
We use the method of Berkowitz et al.\cite{Berkowitz1983} to generate the random 
forces, with each realization given by 
\begin{eqnarray}
\int_{t_{i-1}}^{t_{i}}dt \mathcal{F}(t)&=&\sum_{k=1}^{M}\sqrt{\frac{J(2\pi k/L\Delta t)}{L\Delta t}}\\ \nonumber
&&\times\Bigg[\frac{\zeta_{ak}}{\omega_k}(\sin(\omega_ki\Delta t)-\sin(\omega_k(i-1)\Delta t))\\ \nonumber
&&-\frac{\zeta_{bk}}{\omega_k}(\cos(\omega_ki\Delta t)-\cos(\omega_k(i-1)\Delta t))\Bigg]
\end{eqnarray}
where $\zeta_{ak}$ and $\zeta_{bk}$ are random numbers drawn from independent random gaussian distributions, $\omega_k=2\pi k/(L\delta t)$ 
and $L$ is the number of time steps over which the pseudo-random forces repeat. Therefore $L\delta t$ should be no less than the simulation time.
The spectral density $J(\omega)$ is determined using the memory kernel 
through the Wiener--Khintchine\cite{Kubo1991} theorem,
\begin{eqnarray}
\label{eq:wkt}
J(\omega)&=&4\int_0^\infty dt\Gamma(t)\cos(\omega t)\\ \nonumber
&=&4\left(\Gamma_0+\int_0^\infty dt\Gamma_1(t)\cos(\omega t)\right)
\end{eqnarray}
where $\Gamma(t)=2\Gamma_0\delta(t)+\Gamma_1(t)$. Calculation of the spectral density may need  $\Gamma_1(t)$  at time points finer that those retrieved in Section \ref{subset:TR}, e.g., 
within $(i-1)\Delta t$ and $i\Delta t$, which are obtained through linear interpolation  using $\Gamma_1((i-1)\Delta t)$ and  $\Gamma_1(i\Delta t)$in this work. 


\section{\label{sec:Results}Results}
\subsection{End--product inhibition motif}
To demonstrate the strength of our parameter free projection method, we first apply it to a simple nonlinear chemical network. This network is an end product inhibition motif 
found in metabolic and other biology networks. The reactions are governed by irreversible Michaelis--Menten kinetics,
$$\dot{x}_1=\frac{v_m}{K_m+x_4}-\frac{v_mx_1}{K_m+x_1}+g\xi_1(t),$$
and
\begin{equation}
\dot{x}_i=\frac{v_mx_{i-1}}{K_m+x_{i-1}}-\frac{v_mx_i}{K_m+x_i}+g\xi_i(t) \mbox{      }i=2,3,4
\end{equation}
Values $v_m =1$, $K_m =0.5$, $g =0.005$ are used in the simulations. The  system  initially relaxes to the steady-state. We choose $x_1$ as our species of interest. 
We determine the potential of the mean force defined as $W(x_1)=-\ln(\rho_{ss})$, where  $\rho_{ss}$ is the stationary state distribution of $x_1$. At time 0, the concentration 
of $x_1(0)$ is jumped to $2.0$. The relaxation dynamics $\langle x_1(t )\rangle$, which is defined  as the value of $x_1$ averaged over all trajectories, is recorded for every time step with an interval of $\Delta t=0.1$. In order to compute the memory kernel ($\Gamma_0$, $\Gamma_1(t)$) we first obtain the mean force through histogram counting and fit the entire $W(x_1)$ with $20$ piecewise quadratic functions to facilitate the derivative calculations. Next while recording the $\langle x_1(t )\rangle$ we also record the $\langle\frac{\partial W}{\partial x_1}\rangle$ for every time step determined using the functional form determined using the piecewise quadratic fits.

\begin{figure*}
\centering
\includegraphics{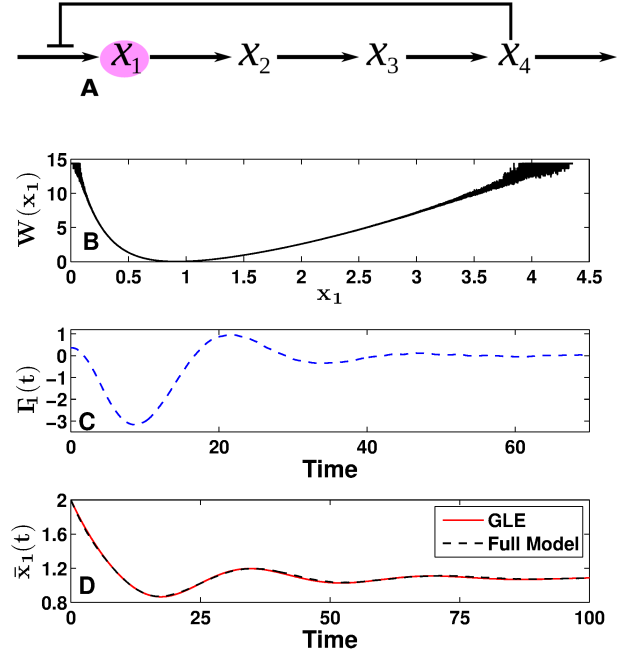}
\caption[Optional caption for list of figures]{(Color Online) Simulations with the end-product inhibition motif. (A) The wiring diagram. $x_1$ is selected as the primary degree of freedom.
(B) The potential of mean force obtained from the stationary state distribution, $W= -\ln(\rho_{ss})$.  (C) The memory kernel, 
$\Gamma_1$ obtained using Eqn. \ref{eq:tikho3} with the Tikhonov regularization penalization factor $b=0.1$. (D) The relaxation dynamics of the species $x_1$ obtained using the GLE compared to the full model simulation results, for initial value $x_1=2.0$. The trajectory for the GLE is averaged over 50000 realizations, and that for the full model is averaged over 300000 realizations.}
\label{fig:jcpex}
\end{figure*}

\begin{figure*}
\includegraphics{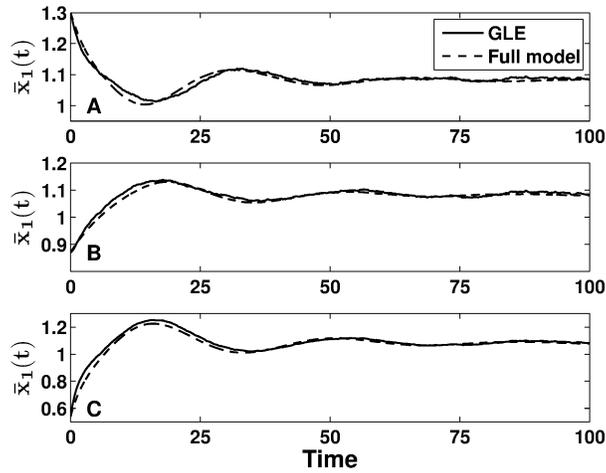}
\caption[Optional caption for list of figures]{Comparison of the Full model vs. GLE results of the $x_1$ relaxation dynamics for different starting points,  
$x_1(0)=1.3 (A)$, $0.87 (B)$ and $0.54 (C)$. The full model simulations are performed over $10^5$ realizations. The same memory kernel as Fig.~\ref{fig:jcpex} is used for the GLE simulations. Each GLE result is performed over $10^4$ realizations.}
\label{fig:otherstpt}
\end{figure*}

Actually this model system has been studied in our earlier work~\cite{Xing2011}, where the memory kernel is obtained by fitting with an ansatz of the function form. Here it is calculated directly using Eqn. \ref{eq:tikho3}. The time independent part of the memory kernel, $\Gamma_0$ obtained  for this system is $28.5$, the time dependent memory kernel, $\Gamma_1(t)$ is shown in 
Fig~\ref{fig:jcpex}. $M$, the number of time steps over which the
pseudo-random forces repeat is chosen to be $1000$ for the GLE whereas the $\Delta t=0.1$.
Both agree well with the previous results~\cite{Xing2011}, but differ in subtle details. As shown in Fig~\ref{fig:jcpex}C, this subtle difference leads to remarkable improvement on the agreement between the GLE result and the full model calculation, compared to the previous work. 
Notice that both $\Gamma_1(t)$ and ${\bar x}_1(t)$ show damped oscillations with similar frequency.  
Furthermore,  with the same memory kernel 
as shown in Fig~\ref{fig:jcpex}, we predict  the relaxation dynamics with different values of $x_1(0)=1.3, 0.87$ and $0.54$. Again, the results in Fig.~\ref{fig:otherstpt} show striking agreements with the full model calculations. 
 In general it is a question to what extent the GLE parameters are transferable from one situation to another one.  The results here give a positive answer.

\subsection{ER-GFR Survival Signaling Switch}
Next we examine a more challenging system shown in Fig \ref{fig:bist}A.
It is a phenomenological model proposed by Tyson et. al. to capture the crosstalk between the growth factor (GF) 
and esptrogen receptor (ER) signaling  in cancer studies~\cite{Tyson2011}. 
All components in GF signaling are lumped into the black box named `GFR', the growth 
factor receptor. Extracellular estrogen, E2-bound esptrogen receptor, ER (ER:E2) inhibits GFR. 
Withdrawn of E2 promotes activation of GFR, which then  activates ERP (phosphorylated ER) and ERP:E2 (E2 bound phosphorylated ER complex). 
ERP facilitates/stabilizes GFR activation. Further details of this model can be found in~\cite{Tyson2011}. 
A remarkable feature of this system is that it can have bistable dynamics.
Appendix~\ref{appendixERGFR} gives the detailed scheme for the full model and the GLE simulations. 
We use chemical Langevin equations to simulate the stochastic dynamics.  Table~\ref{tab:t1} in Appendix \ref{appendixERGFR} gives the model parameters. 

Figure \ref{fig:bist}B shows that this bistable system has a double-well shaped potential of mean force, with the more stable (left) state corresponding to the low GFR activity state. 
To generate the data for reconstructing the GLE, we initialize the system within the right well corresponding to the high GFR activity state.  Fig.~\ref{fig:bist}C gives the obtained memory kernel $\Gamma_1(t)$. It seems to reach zero at $t\sim 5\times10^4$ mins. As 
our first attempt, we hence set $\Gamma_1(t)=0$ for $t>5\times10^4$. However,   the blue curve in Fig.~\ref{fig:bist}D shows that the GLE result  does not
agree well with the full model simulation results except at an early stage and the long time behavior. For the latter it is because the long time steady state behavior is governed by the potential of mean force. For the former the early stage dynamics is governed by $\Gamma_0$. The discrepancy in the middle stage suggests that $\Gamma_1$ is not sufficiently long. Physically, we think that the relaxation process involves the fast intrawell dynamics and the slow inter-well dynamics. In this case the inter-well transition dynamics is beyond a simple Markovian process, but shows correlation among transitions.  
Correspondingly, the memory kernel $\Gamma_1$ shows a rather long tail (Fig. \ref{fig:bist}C), and a biphasic relaxation dynamics for ${\bar x}(t)$ (Fig. \ref{fig:bist}D). 
Indeed using the evaluated $\Gamma_1(t)$ up to $t=10^8$ mins, we obtain an excellent agreement between the GLE (red curve in Fig.~\ref{fig:bist}D) and the  full model simulations. 

\begin{figure*}
\includegraphics{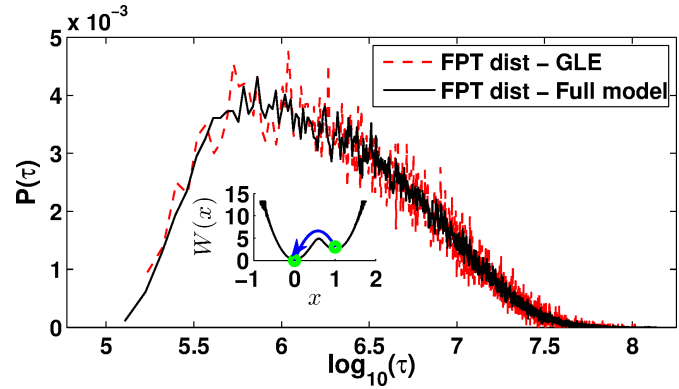}
\caption[Optional caption for list of figures]{(Color Online) Distributions ($P(\tau)$) of the first passage time ($\tau$) from the shallow well 
(near $A_{GFR}=x=1$) to the deep well (near $A_{GFR}=x=0$) using full model and 
the generalized langevin equation for the ER- GFR switch. Note that the x-axis is the natural 
log of the first passage time.}
\label{fig:fpt}
\end{figure*}

Next, we use the reconstructed GLE to predict the first passage time distribution for the transition from the right well (high GFR) to the left well (low GFR). 
Fig. \ref{fig:fpt} shows excellent agreement between the GLE and full model results. It is a highly nontrivial achievement to reproduce the whole distribution, not just a few moments, such as the mean first passage time.


\section{\label{sec:Discussions} Discussions}
Reconstruction of the governing equations is an important step towards understanding a dynamical system. The Zwanzig-Mori projection method provides a rigorous theoretical formalism.
In this work, we present a numerical procedure to reconstruct the GLEs from the data. Numerical tests on two model systems show that the accuracy of the algorithm is encouraging.

Eqn. \ref{eqn:GLE_formal} is a mathematically exact solution. It is equivalent to but mathematically more involved than the original dynamic equations.  The practical usage of the projection approach lies in the assumption that the effect of the implicitly treated secondary degrees of freedom can be well replaced by the potential of mean force, the memory kernel with some simple function forms, and the related noises with certain statistical properties. Therefore one expects that the GLEs after such  approximation in the form of Eqn. \ref{eqn:GLE_phe} work best for systems coupled to many degrees of freedom. Einstein shows a classical example that a single drag coefficient and a corresponding white noise term can well describe the dynamics of a Brownian particle influenced by its interaction with Avogadro number of solvent molecules. Then it is out of surprise that for both of the two models, which have only a small number of degrees of freedom, the GLE results and the full model simulations agree remarkably well. It may be partly due to the fact that the full system dynamics is governed by stochastic differential equations, and the stochastic noise can be viewed as generated by a bath system with infinite number of degrees of  freedom~\cite{Xing2010}.  It remains to examine to what extent one can use a coordinate-independent memory kernel form to describe a dynamic system.

Most molecular systems are well modeled by some simple memory kernel forms such as an exponentially decayed function. Recent single molecule studies reveal a power-law form for describing intramolecular fluctuations~\cite{Min2005}. Studies on the two model systems in this work show that the function form can be more complex for systems kept out of equilibrium. Can one find a set of common function forms corresponding to systems with different dynamic behaviors? Let's ask the problem in an alternative way. In two or higher dimensions, the GLEs in Eqn. \ref{eqn:GLE_phe} share similar form as those obtained for Hamiltonian systems relaxing to equilibrium, except that the matrix $\Gamma_0$ is not symmetric. For a one-dimensional GLE, however, the form is the same. Then can one tell whether it describes a closed or open system from the equation itself?   

In this work we perform numerical simulations with the stochastic GLEs to demonstrate their accuracy. 
Simulating GLEs with colored noise is computationally expensive, especially if the noise spectrum has a long tail. In practical applications, in general there is no need to perform such full stochastic simulations. Instead one may just need to calculate the first few moments of the distribution, or work with the corresponding Fokker-Planck equation.

Our numerical studies also reveal that the quality of a reconstructed GLE is very sensitive to the accuracy of the potential of mean force, which then requires well converged sampling. Some systematic studies and statistical tools are needed to improve the accuracy of reconstruction with less sufficient data. In our current algorithm, we first obtain the potential of mean force, and then the memory kernel that is affected by the form of the former. It may be desirable to have a procedure to reconstruct the two simultaneously.  

Model 2 examined in this work shows a very long memory kernel, which includes both intra- and inter-well dynamics. One possible reason  is that $A_{GFR}$ is not a good choice for the reaction coordinate, and the dynamics along the orthogonal degrees of freedom is comparable or even slower than that along $A_{GFR}$. Lange and Grubmuller demonstrated how to reconstruct a GLE along a nonlinear collective coordinate~\cite{Lange2006}. One may follow similar procedure to examine whether a short memory kernel can be obtained, and whether the dynamics between the two states can be approximated as Markovian processes.

In this work we focus on model reduction. Since the data needed for reconstruction is in the form of time series, the procedure can be used as a way of nonlinear time series analysis~\cite{Kantz_book, Schmitt2006}. The latter is an active and under-developed area, with a main difficulty lying in how to construct the nonlinear function form. In the GLE formalism, however, the nonlinearity is  given by the potential of mean force that is automatically obtained from data.  Therefore in principle the difficulty of selecting the nonlinear function form does not exist.

In conclusion, in this work we demonstrate the practical feasibility and accuracy of using the GLEs to model the dynamics of a general dynamic system, \textit{i.e.}, with or without detailed balance constraint, in reduced dimension. Further developments are necessary to test the formalism in different systems and make the algorithm practically useful on studying complex systems.

\section{Acknowledgements} We thank Steve Press\'e for suggesting the Tikhonov regularization method. This work is supported by National Science Foundation (DMS-0969417), and the Institute for Critical Technology and Applied Science of Virginia Tech. We also thank the Athena cluster at the Advanced Research Computing facility at Virginia Tech for providing computational resources.
\appendix
\section{\label{appendixAnalytic}An Analytical Example}
Here we consider a system-bath Hamiltonian,
\begin{equation}
H=\frac{p^2}{2}+\frac{x^2}{2}+\frac{b}{4}x^4 + \sum_j\left\{\frac{p_j^2}{2}+\frac{\omega_j^2}{2}\left(q_j-\frac{\gamma_j}{\omega_j^2 }x \right)^2\right\}
\end{equation}
Zwanzig discussed a nonlinear GLE for the system coordinates $\{x, p\}$  obtained by directly solving the equations of motion~\cite{Zwanzig1973, Zwanzig2001}, 
\begin{eqnarray}
\frac{dx(t)}{dt} &=& p(t)\nonumber\\
\frac{dp(t)}{dt} &=& -x(t)-bx(t)^3 \nonumber\\
&& -\int_0^t dsK_N(s)p(t-s)+F_N(t)\label{eqn:GLE_exact}
\end{eqnarray}
The memory kernel and the random force terms are given by,
\begin{eqnarray}
K_N(t) &=& \sum_j \frac{\gamma_j^2}{\omega_j^2}\cos(\omega_j t) \label{eqn:memory_exact}\\
F_N(t)  &=& \sum_j \gamma_j p_j(0) \frac{\sin \omega_j t} {\omega_j} \nonumber\\
&& + \sum_j \gamma_j \left(q_j(0)-\frac{\gamma_j}{\omega_j^2}x(0)\right)\cos \omega_j t\label{eqn:randforce_exact}
\end{eqnarray}
with the fluctuation-dissipation relation,
\begin{eqnarray}
<F_N(t) F_N(t')>_0 &=& k_BTK_N(t-t').
\end{eqnarray}
The average is over an equilibrium heat bath with the system constrained at $\{x(0),p(0)\}$.
By projecting to the Hilbert space $(x,v)$ with Mori's procedure, one can also obtain a linearized GLE~\cite{Zwanzig2001},
 \begin{eqnarray}
\frac{dx(t)}{dt} &=& v(t)\nonumber\\
\frac{dp(t)}{dt} &=& -\omega_0^2 x(t)-\int_0^t dsK_L(s)p(t-s)+F_L(t)
\end{eqnarray}
Where $\omega_0^2=k_BT/<x^2>$, and the the random force and memory kernel terms are also related by the fluctuation-dissipation relation
\begin{eqnarray}
<F_L(t) F_L(t')> &=& k_BTK_L(t-t'). 
\end{eqnarray}
However in this case, the average is over the unconstrained thermal equilibrium distribution. Effects of the nonlinear term $-bx(t)^3$ are contained in the renormalized
coefficent $\omega_0^2$, the memory kernel, and the random force terms.

In the following discussions, we will generalize the projection procedure of Mori by choosing a basis set $\{x,x^3,v\}$. Functions 
with even powers of $x$ makes no contribution to the projection ($(Lp,x^{2n})=0, n=1,2,\dots$). Therefore the lowest nolinear basis function is $x^3$.

First,
\begin{eqnarray}
Lx = p, Lx^3 = 3px^2\nonumber\\
Lp = -x-bx^3-\sum_i \gamma_i\left(\frac{\gamma_i}{\omega_i^2}x-q_i \right)
\end{eqnarray}
 let's calculate the normalization matrix,
\begin{eqnarray}
A^{-1}&=&\left(\begin{array} {ccc}
          <x^2> & <x^4>   & <xp>\\
          <x^4> & <x^6>   & <x^3p>\\
          <xp>   & <x^3p> & <p^2>
          \end{array}    
\right)^{-1}\\ \nonumber
&=& \left(\begin{array} {ccc}
          <x^6>/h & -<x^4>/h   & 0\\
          -<x^4>/h & <x^2>/h   & 0\\
          0  & 0 & <p^2>^{-1}
          \end{array}\right) 
\end{eqnarray}
Where $h=<x^2><x^6>-<x^4>^2$.
The memory function and the random force in the equation of motion of $x$ vanish, which can be seen from,
\begin{eqnarray}
Lx &=&  \left(\begin{array} {ccc}
          (Lx,x) & (Lx,x^3)   & (Lx,p)
          \end{array}\right)\cdot A^{-1} \nonumber\\
          &=& p   
\end{eqnarray}
One has,
\begin{eqnarray}
(Lp,x^n) &=& -\frac{1}{\int \exp(-\beta H)dx}\int x^n \frac{\partial H }{\partial x} \exp(-\beta H)dx\nonumber\\
&=& \frac{k_BT}{\int \exp(-\beta H)dx}\int x^n \frac{\partial}{\partial x} \exp(-\beta H)dx\nonumber\\
&=& -\frac{k_BT}{\int \exp(-\beta H)dx}\int nx^{n-1} \exp(-\beta H)dx\nonumber\\
&=&   -n k_BT<x^{n-1}>\\
(Lp,x) &=& -k_BT= -\omega_0^2<x^2>
\end{eqnarray}
However, one also has,
\begin{eqnarray}
<Lp,x^n> &=& -<x^{n+1}> -b<x^{n+3}>  \nonumber\\
&&- \sum_i \gamma_i <x^n (\frac{\gamma_i}{\omega_i^2}x - q_i)>\nonumber\\
&=& -<x^{n+1}> -b<x^{n+3}> 
\end{eqnarray}
Therefore,
\begin{eqnarray}
<x^4> &=& \frac{1}{b}(\omega_0^2-1)<x^2>\\
<x^6>  &=&  \frac{3}{b} \omega_0^2<x^2><x^2> \nonumber\\
&& -\frac{1}{b^2}(\omega_0^2-1)<x^2>
\end{eqnarray}
  Then,
\begin{eqnarray}
 PL p(t) = -x-bx^3
\end{eqnarray}

One can easily show that the random force (through $dF/dt = (1-P)LF$) and memory kernel (through Eqn. \ref{eqn:memory_general}) 
terms are the same as those given in Eqs. \ref{eqn:memory_exact} and \ref{eqn:randforce_exact} , although in general here the average 
perform in Eqn. \ref{eqn:memory_general} is over the unconstrained thermal equilibrium distribution. Therefore with the Mori projection 
procedure we recover Eqs. \ref{eqn:GLE_exact}, \ref{eqn:memory_exact}, \ref{eqn:randforce_exact}  obtained by exact integration. 
Following similar procedure, one can show that further expanding the basis functions to include higher orders of $x^n$ does not change 
the projected equation form. The above results can also be obtained by applying Eqs. \ref{eqn:GLE_formal}, \ref{eqn:collective_c},
\ref{eqn:collective_cdot} directly.
\begin{widetext}
\begin{table*}[htbp]
\centering
\caption{Parameters table for the full model simulation of the ER-GFR signaling switch.}
\begin{tabular}{@{} l|c|l @{}} 
      \hline
\textbf{Parameter} &\textbf{Description}& \textbf{Value} \\
      \hline
$k_{dsER:E2}$& Dissociation of $ER:E2$ and $ERP:E2$ & $0.001 min^{-1}$\\ 
$k_{asER:E2}$ & Association of $ER:E2$ and $ERP:E2$ & $0.001 nM^{-1} min ^{-1}$\\
$k_{dpERP}$ & Dephosphorylation of $ERP$ and $ERP:E2$ & $0.001 min ^{-1}$\\  
$J_{dpERP}$& Michaelis constant for dephosphorylation &$1.8 nM$\\  
$k_{pER}$ & Phosphorylation of $ER$ and $ER:E2$ &$0.0011 min ^{-1}$\\  
$J_{pER}$& Michaelis constant for phosphorylation & $3 nM$\\  
$k_{sER}$& Production rate of $ER$ &$0.001 nM min^{-1}$\\
$k_{dER}$ & Degradation rate of $ER$ & $10^{-5} min^{-1}$\\ 
$[E2]$& Extracellular concentration of estrogen &$0.003 nM$\\ 
$\gamma$& Time-scale for $GFR$ activation & $10^{-6} min^{-1}$\\  
$\sigma$& Sigmoidicity of $GFR$ response function & $1.9$\\  
$\omega_0$& Basal inactivation of $GFR$&  $-0.83$\\  
$\omega_1$& GFR inactivation by $ER:E2$ & $-0.5$ \\ 
$\omega_2$,$\omega_3$,$\omega_4$& $GFR$ activation by $ERP:E2$, $ERP$ and $GFR$ & $0.5, 0.5, 1.1\times10^{-3}$\\
$\rho_1$,$\rho_2$,$\rho_3$,$\rho_4$,$\rho_5$&Amplitudes of noise for $ER$, $ERP$, $ER:E2$, $ERP:E2$ and $A_{GFR}$ & $0.05, 0, 0, 0, 0.15$\\
      \hline
\end{tabular}
\label{tab:t1}
\end{table*}
\end{widetext}
\section{\label{appendixERGFR}Implementing the stochastic full model simulation for ER-GFR signaling switch and reconstructing the GLE}
The rate equations for the ER-GFR signaling switch are taken from~\cite{Tyson2011} and are listed below;
\begin{widetext}
\begin{eqnarray}\nonumber
&&\frac{d[ER]}{dt}=k_{sER}-k_{dER}[ER]+k_{dsER:E2}[ER:E2]-k_{asER:E2}[ER][E2]+\frac{k_{dpERP}[ERP]}{J_{dpERP}+[ERP]}-\frac{k_{pER}A_{GFR}[ER]}{J_{pER}+[ER]}\\ \nonumber
&&\frac{d[ERP]}{dt}=-k_{dER}[ERP]-k_{dsER:E2}[ERP:E2]-k_{asER:E2}[ERP][E2]-\frac{k_{dpERP}[ERP]}{J_{dpERP}+[ERP]}+\frac{k_{pER}A_{GFR}[ER]}{J_{pER}+[ER]}\\ \nonumber
&&\frac{d[ER:E2]}{dt}=-k_{dER}[ER:E2]-k_{dsER:E2}[ER:E2]+k_{asER:E2}[ER][E2]+\frac{k_{dpERP}[ERP:E2]}{J_{dpERP}+[ERP:E2]}\\ \nonumber
&&\mbox{              }\mbox{              }\mbox{              }\mbox{              }\mbox{              }\mbox{              }\mbox{              }\mbox{              }\mbox{              }\mbox{              }\mbox{              }\mbox{              }\mbox{              }\mbox{              }\mbox{              }\mbox{              }\mbox{              }\mbox{              }\mbox{              }\mbox{              }\mbox{              }\mbox{              }-\frac{k_{pER}A_{GFR}[ER:E2]}{J_{pER}+[ER:E2]}\\ \nonumber
&&\frac{d[ERP:E2]}{dt}=-k_{dER}[ERP:E2]-k_{dsER:E2}[ERP:E2]+k_{asER:E2}[ERP][E2]-\frac{k_{dpERP}[ERP:E2]}{J_{dpERP}+[ERP:E2]}\\ \nonumber
&&\mbox{              }\mbox{              }\mbox{              }\mbox{              }\mbox{              }\mbox{              }\mbox{              }\mbox{              }\mbox{              }\mbox{              }\mbox{              }\mbox{              }\mbox{              }\mbox{              }\mbox{              }\mbox{              }\mbox{              }\mbox{              }\mbox{              }\mbox{              }\mbox{              }\mbox{              }+\frac{k_{pER}A_{GFR}[ER:E2]}{J_{pER}+[ER:E2]}\\ 
&&\frac{dA_{GFR}}{dt}=\gamma \left[\left({1+e^{-\sigma\left(\omega_0+\omega_1[ER:E2]+\omega_2[ERP:E2]+\omega_3[ERP]+\omega_4[GFR]\right)}}\right)^{-1}-A_{GFR}\right]
\label{eq:rate}
\end{eqnarray}
\end{widetext}
where, $[ER]$ is the concentration of unbound estrogen receptor; $[ERP]$, the concentration 
of unbound phosphorylated estrogen receptor;  $[ER:E2]$, the concentration of estrogen receptor 
bound to estrogen ($[E2]$);  $[ERP:E2]$, the concentration of phosphorylated estrogen receptor 
bound to estrogen and $A_{GFR}$ is defined as the activity of growth factor receptor defined as 
$log_{10}[GFR]$. Table~\ref{tab:t1} lists all the model parameters.
\\ \\
Following Tyson \textit{et al},  we convert the deterministic rate equations Eqn.~\ref{eq:rate} to chemical 
Langevin equations as follows. 
Specifically we add nonzero noise terms to the equations of $ER$ and $A_{GFR}$,
\begin{equation}
\frac{dX_i}{dt}=\lambda_i\cdot\left(S_i-X_i\right)+R_i(X_1, \cdot\cdot\cdot, X_i,\cdot\cdot\cdot, X_n)+\sqrt{2\lambda_i}\rho_i\xi_i(t)
\label{eq:ergfrcle2}
\end{equation}
where $\lambda_i$ determines the time scale of synthesis and degradation reaction of species $X_i$, $S_i$ represents 
the steady state level of species $X_i$,  $R_i$ indicates the rate of all other reactions affecting $X_i$, $\rho_i=\sqrt{\langle\left(S_i-X_i\right)^2\rangle_{eq}}$, and $\xi_i$ is a Gaussian noise term. Only fluctuations  arising from the processes of protein synthesis and degradation are added to the model; noise terms arising from the fast 
association disassociation and phosphorylation--dephosphorylation reactions are ignored. 
In our simulations, a hard reflecting barrier is imposed at 0 for all chemical species (but not for $A_{GFR}$), so that only positive concentrations 
are involved. For $A_{GFR}=log_{10}[GFR]$, a negative value 
corresponds to concentration of GFR less than 1, so there is no scope of negative concentration of $GFR$. 
\\ \\

\begin{figure*}
\centering
\includegraphics{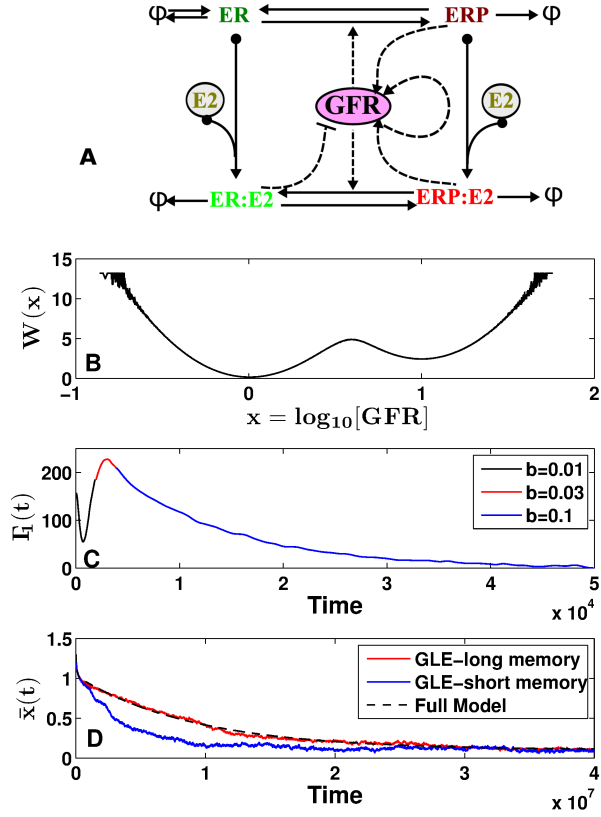}
\caption{(Color Online)  Model studies of the ER-GFR signaling network. (A) The wiring diagram adapted from~\cite{Tyson2011}. Each solid line starts with the reactant(s), and ends with the product(s). For a reaction with multiple reactants, each reactant is indicated with a filled solid circle. Each dashed line represents the influence, 
with a point arrow end for activation, and a blunt end for inhibition. The symbol $\phi$ means degradation.
(B) The potential obtained from the stationary state distribution, $W\sim -\ln(\rho_{ss})$. (C) The memory kernel, 
$\Gamma_1$ obtained using the full model simulation and smoothened using Tikhonov regularization with the penalization factor, 
$b=0.01$ (black) -- early behavior, $0.03$ (red) -- intermediate behavior and $0.1$ (blue) -- 
late behavior, and $\Gamma_0\simeq1.0\times 10^6$. 
(D) Comparison of GLE with the full model simulation using the time evolution of $x=A_{GFR}=log_{10}[GFR]$. The initial value of $x=1.3$. The black curve is the full model simulation averaged over $10^6$ realizations. The red and the blue curves are generated using the GLE with memory kernel for $10^8$ and $5\times10^4$ time steps, respectively, each averaged over $10^3$ realizations. The better agreement of the red curve justifies the importance of the long tail of the memory kernel.}
\label{fig:bist}
\end{figure*}

For reconstruction of the GLE, we choose $A_{GFR}$ as our species of interest. The chemical Langevin equation, described 
by Eqns.~\ref{eq:rate} and~\ref{eq:ergfrcle2} with parameters from the Table~\ref{tab:t1}, is simulated for $2\times10^6 mins$ 
with a time step of $1000$ mins using initial concentration of $[E2] = 0 nM$, $[ER]=19.04 nM$, $[ERP]=3.15 nM$, 
$[ER:E2]=[ERP:E2]=0$, $A_{GFR}=0.936$. Thereafter the stationary state distribution, $\rho_{ss}$ is obtained 
by sampling $1.4\times10^7$ points. The mean force potential, Fig~\ref{fig:bist}B, is obtained using the relation $W=-\ln(\rho_{ss})$.

To study the dynamics, $A_{GFR}$ is dragged out of stationary state and set to $1.3$, while the concentration of the 
rest of the species are sampled from the original stationary distribution. To determine the memory kernel 
($\Gamma_0$ and $\Gamma_1(t)$) the mean potential is fit with $20$ piecewise quadratic functions. Using the functional 
form of the mean force obtained via the quadratic fits and Eqn.~\ref{eq:memk} the memory kernel is determined for $\Delta t=50$. 
We would like to point out that a large value of the penalizing factor in Tikhonov regularization results in inaccurate estimate of the $\Gamma_1(t)$ while a small 
value accumulates error with time. Hence four different values of $b$ are used  for different intervals {\it viz.} $b=0.01, 0.03, 0.1, 3$ 
for $t\in(0,2000)$, $(2\times10^3,4\times10^3)$,  $(4\times10^3,5\times10^4)$, and beyond  $5\times10^4$ mins, respectively. 
Although $b=3$ is a large penalization, the general signature {\it i.e.} a slow relaxation of the long tail, $t>5\times10^4$ mins, of the memory kernel remains 
intact. For the spectral density calculations, $M$, the number of time steps over which the pseudo-random forces repeat is chosen 
to be $1000$ for the GLE reconstruction with short memory kernel and $2\times10^6$ for the long one, corresponding to the trajectories 
shown in blue and red curves respectively in Fig~\ref{fig:bist}D, with time step $\Delta t=50$. 
\\ \\
Since these simulations are computationally expensive, the codes are parallelized using {\it mpirun} in 
{\it athena} cluster at Virginia Tech advanced research computing facility on 8 cpu nodes with 16 processors 
per node.


\bibliographystyle{apsrev}

\end{document}